\documentclass{article}
\usepackage{amsmath,amssymb}

\usepackage{geometry}
\title{Ultrahigh energy cosmic rays from  a nearby \\ extragalactic source in the diffusive regime}
\author{Silvia Mollerach and Esteban Roulet\\
Centro At\'omico Bariloche, Comisi\'on Nacional de Energ\'\i a At\'omica\\ 
Consejo Nacional de Investigaciones Cient\'\i ficas y T\'ecnicas (CONICET)\\ 
Av. Bustillo 9500, R8402AGP, Bariloche, Argentina}
\date{}

\usepackage{graphicx}

\begin{document}

\maketitle
\begin{abstract}
We study the effects that the diffusion of the cosmic rays in the magnetic field of the Local Supercluster can have on the spectrum of a nearby extragalactic source at ultrahigh energies. We find that the strong enhancement of the flux below the energy at which the transition between the diffusive and quasirectilinear regimes takes place, as well as the suppression at lower energies associated with a finite source age, can help to explain the observed features of the cosmic-ray spectrum and the composition.  Scenarios are discussed in which a nearby extragalactic source  with mixed composition and rigidity-dependent spectrum accounts for most of the observed cosmic rays at  energies above a few EeV while the rest of the extragalactic sources lead to a diffuse flux that dominates at lower energies and down to $\sim 0.1$~EeV. The nearby source can also naturally account for the dipolar anisotropy measurements above 4~EeV, and these measurements can also help to constrain its evolution with redshift.
\end{abstract}

\section{Introduction}
The origin and nature of the ultrahigh energy cosmic rays (UHECRs), i.e. those with energies above one EeV, is still unknown in spite of many decades of experimental and theoretical studies.  There are two main features in their spectrum: the hardening observed at $\sim 5$~EeV, known as the `ankle', and the  suppression observed above $\sim 40$~EeV \cite{augerspec,taspec}.  On the other hand, the measurements of the composition suggest that at EeV energies the cosmic rays are predominantly light, consisting mostly of H and He, while above few EeV their average mass becomes increasingly heavier \cite{augercomp}. 

Scenarios that have been proposed for the UHECRs include ones with pure proton sources reaching maximum energies in excess of a few hundred EeV, in which the high-energy suppression arises from the Greisen-Zatsepin-Kuzmin effect \cite{gzk}, i.e. from the attenuation that they suffer by photopion production off  cosmic microwave background (CMB) photons. The ankle feature has been associated in this case to a propagation effect related to the threshold for pair production with the same background photons, in the so-called dip scenario \cite{dip}. Although in such scenario the main features of the spectrum  can be naturally accounted for, the observed composition change is not explained and also the source properties that are  required in order to accelerate protons to such high energies are quite demanding.  The spectral hardening at the ankle has alternatively been associated with the transition between a steeply falling Galactic component at lower energies and an emerging harder extragalactic component. The main difficulty with this kind of scenario is that, given the relatively light composition inferred at EeV energies, if those CRs were of Galactic origin a strong anisotropy towards the Galactic center and the Galactic plane would be expected, something that is not observed \cite{augergc}.
Moreover, there are no natural Galactic source candidates to accelerate protons to few EeV energies and the transition between predominantly Galactic and extragalactic sources probably takes place instead near 0.1~EeV. 
When the spectrum and composition  measurements from the Pierre Auger Observatory are fitted in terms of homogeneously distributed extragalactic sources with power-law spectra \cite{combfit},   a low cutoff value for the maximum cosmic-ray rigidities, $E/Z \lesssim 5$~EeV, tends to be preferred so as to lead to an increase in the average mass above the ankle energy. Moreover,  below this cutoff value very hard source spectra, $\phi(E)\propto E^{-\gamma}$  with $\gamma\simeq 1$, are required in order that the heavier components that dominate the flux at the highest energies  be sufficiently suppressed near the ankle energy.
Including a turbulent extragalactic magnetic field and accounting for the finite density of sources can give rise to a low-energy magnetic horizon effect that could make the observations compatible with a larger spectral index  \cite{mo13,wi18},  closer to the values expected from diffusive shock acceleration (which are $\gamma\simeq 2$--2.4).
Still other scenarios rely on extragalactic sources accelerating  heavy nuclei which remain magnetically confined for long times around their sources and, interacting with the IR radiation present in those environments, photodisintegrate to produce a large number of secondary nucleons that could account for the light composition present below the ankle \cite{un15,gl15}.

We want here to propose an alternative scenario in which the dominant contribution to the observed CR fluxes at energies  above few EeV arises from a powerful nearby extragalactic source, which may have been in a stage of enhanced emission since relatively recent times (due to a galaxy merging event, an enhanced accretion rate in an active galactic nucleus, a strong burst of star formation, etc.). The spectrum at lower energies and down to $\sim 0.1$~EeV could instead be dominated by the diffuse contribution from  the large number of extragalactic CR sources that are present up to high redshifts. A crucial ingredient for this scenario is the diffusion of the CRs from the nearby source in the turbulent intergalactic magnetic field, which is expected to be sizable  in the Local Supercluster region.  This should enhance significantly, by up to more than an order of magnitude, the CR density due to the nearby source with respect to the expectations in the case of rectilinear propagation. The source will be considered to have a mixed composition, with rigidity-dependent spectra. If the change between the regimes of diffusive and quasirectilinear propagation takes place at rigidities $E/Z\simeq 10$--30~EeV, the ankle can naturally be explained as due to the suppression of the light components resulting from the energy dependence of the diffusion enhancement, without the need to invoke any source cutoff.  Moreover, the apparently hard spectra of the individual mass components that is inferred from the observations
can be naturally related to a magnetic horizon effect if the nearby source is relatively young, so that at low energies it would take longer than the age of the source for the CRs to arrive at the Earth. Finally, a nearby source would give rise to an anisotropy in the cosmic-ray flux that can also explain the observed dipolar amplitudes \cite{dipolescience, dipoleapj}.

The diffusive CR propagation from a source in the Local Supercluster, such as Virgo or Cen A, was studied previously in \cite{gi80,be90,le99,bl99,fa00}, focusing in proton sources. The consideration of a nearby source in those cases allowed to mitigate the spectral suppression due to interactions with the CMB, and the diffusion helped to steepen its spectrum and to reduce the anisotropies associated with that source.

\section{Turbulent magnetic fields and diffusive propagation}

 Only a few observational constraints exist on the extragalactic magnetic fields, making it difficult to construct a model for them (for a review see \cite{han17}). In galaxy clusters, they are probed through the measurement of the Faraday rotation effect on the light of embedded and background galaxies and also through radio emission from diffuse synchrotron sources, both near their centers and in their periphery.
These large-scale magnetic fields have measured amplitudes that range from a few up to tens of $\mu$G in the cluster central regions \cite{fe2012}. This suggests that significant large-scale magnetic fields should also be present in cosmic structure filaments and sheets, with strengths that could range from nG up to $\mu$G, although measurements of them are still lacking. The magnetic fields are  expected to have smaller strengths in the void regions, and typical bounds on the magnetic fields in unclustered regions are $B<1$--10~nG.

 We will be interested here in the study of the propagation of the CRs from a nearby extragalactic source within the Local Supercluster region
to which  the Milky Way belongs and which includes, besides the Local Group,  the Virgo, Leo, Ursa Major, Draco and other clusters, extending for about 30~Mpc. 
The presence of a large magnetic field in the Local Supercluster, with strength possibly as large as 0.3 to 2~$\mu$G, has been suggested from the observed rotation measure of polarized background sources \cite{vallee}.
We will consider a simplified description of the intergalactic magnetic field in this region, describing it as a turbulent isotropic field with root mean square strength $B=\sqrt{\langle B^2(x)\rangle}$, which could take a value in the range from few tens of nG up to few hundred~nG, and having a coherence length with typical values of order $l_{\rm c}\sim 0.01$--1~Mpc. The distribution of the magnetic energy density $w$ on different length scales is described by adopting a power law in Fourier space, $w(k)\propto k^{-m}$. In particular, we will consider the case of a Kolmogorov spectrum of turbulence for which $m=5/3$. Note that in this case the coherence length is related to the maximum scale of the turbulence $L_{\rm max}$ through $l_{\rm c}\simeq 0.2 L_{\rm max}$ \cite{ha02}.

For charged particles propagating in a turbulent magnetic field, an effective Larmor radius can be introduced as
\begin{equation}
r_{\rm L}=\frac{E}{ZeB}\simeq 1.1 \frac{E/{\rm EeV}}{Z\,B/{\rm nG}}\,{\rm Mpc},
\end{equation}
with $Ze$ the particle charge. A relevant quantity to characterize the particle diffusion is the critical energy $E_{\rm c}$, defined such that $r_{\rm L}(E_{\rm c})=l_{\rm c}$. It is given by

\begin{equation}
E_{\rm c}=Z e B l_{\rm c} \simeq 0.9 Z\,\frac{B}{\rm nG}\frac{l_{\rm c}}{\rm Mpc}\,{\rm EeV}.
\end{equation}
For energies below $E_{\rm c}$, the regime of resonant diffusion takes place, in which particles experience large deflections induced by their interactions with the $B$ field modes with scales comparable to the Larmor radius. For energies above $E_{\rm c}$, the non-resonant diffusion regime holds, in which the deflections after traversing a distance $l_{\rm c}$ are small, typically of order $\delta\simeq l_{\rm c}/r_{\rm L}$.
 
  From the results of extensive numerical simulations of the propagation of protons, a fit to the diffusion coefficient $D$ as a function of the energy was obtained in \cite{hmr14}. It is given by
\begin{equation}
D(E) \simeq \frac{c}{3}l_{\rm c}\left[ 4\left(\frac{E}{E_{\rm c}}\right)^2 + a_I\left(\frac{E}{E_{\rm c}}\right) + a_L\left(\frac{E}{E_{\rm c}}\right)^{2-m}\right].
\label{D(E).eq}
\end{equation}
 For a Kolmogorov spectrum for the turbulent magnetic field ($m=5/3$), the resulting coefficients are $a_L\approx  0.23$ and $a_I\approx  0.9$.
The diffusion length is defined as $l_D\equiv 3D/c$ and corresponds to the distance after which the total deflection of the particles is about 1~rad. We see from eq.~(\ref{D(E).eq}) that for $E\ll 0.1 E_{\rm c}$ it is given by $l_D\simeq a_L l_{\rm c}(E/E_{\rm c})^{2-m}$. On the other hand, for $E\gg 0.2 E_{\rm c}$ the diffusion length  is $l_D\simeq 4~ l_{\rm c}(E/E_{\rm c})^2$, since in this regime one needs to traverse $N\simeq l_D/l_{\rm c}$ coherent domains to have a total deflection $\delta\simeq 1$~rad (where $\delta\simeq \sqrt{N}(l_{\rm c}/r_{\rm L})$ results from the random angular diffusion of the CR trajectory).

Spatial diffusion of the CR particles takes place whenever the distance to the source $r_{\rm s}$ is much larger than $l_D$. However, at sufficiently large energies the quasirectilinear regime would eventually be reached when  $l_D$ becomes much larger than $r_{\rm s}$. This happens for $E>E_{\rm rect}\equiv E_{\rm c}\sqrt{r_{\rm s}/l_{\rm c}}$ (where we assumed that $E_{\rm rect}>E_{\rm c}$ so that $D\propto E^2$, which is indeed the case if $r_{\rm s}\gg l_{\rm c}$, and note that with this definition one has $l_D(E_{\rm rect})\simeq 2r_s$). In this case, the root mean square deflection of the particles arriving from the source will be less than 1~rad and hence  only some angular diffusion would take place, making the image of the source to appear  blurred.  
For a steady source and in the simplified case in which energy losses are neglected,
it has been shown in \cite{hmr14} that the average cosine of the deflection $\theta$ between the cosmic-ray arrival direction and the line of sight to the source is accurately fitted by
\begin{equation}
\langle\cos\theta\rangle=\frac{1}{3R}\left[1-\exp\left(-3R-\frac{7}{2}R^2\right)\right] \equiv C(R),
\label{cos}
\end{equation}
 with $R\equiv r_{\rm s}/l_D=cr_{\rm s}/3D$.

\section{Spectrum from one source} 

Let us first consider the simple case of a steady source, i.e. one that had a constant intensity  for a very long time, neglecting the energy losses and cosmological evolution effects. The spatial density of the particles from the source will reach an asymptotic stationary regime in which it does not depend on time.  In this case, the flux through any sphere around the source has to be the same and, due to the spherical symmetry, one necessarily has that
\begin{equation}
n(E,r_{\rm s})  4\pi r_{\rm s}^2 c \langle \cos \theta (E, r_{\rm s}) \rangle = Q(E) ,
\label{ner}
\end{equation}
where $n(E,r_{\rm s}) $ denotes the density of the particles at the observer's location, $c$ is  the speed of light and $Q(E)$ is the emissivity of the source (both $n$ and $Q$ are differential in energy, i.e. they reflect the  spectrum of the source). 

For small distances, for which the propagation is nearly rectilinear and  $\langle \cos \theta \rangle \simeq 1$, the density decreases as $r_{\rm s}^{-2}$ and the spectrum of the observed particles coincides with the emitted one. 
On the other hand, when the diffusion length is  smaller than the distance to the source (which is the case for $E<E_{\rm rect}$), there is an enhancement in the density of the particles that is inversely proportional to $\langle \cos \theta \rangle$. From eq.~(\ref{cos}) one can see that in the limit $ l_D\ll r_{\rm s}$ one has that $\langle \cos \theta \rangle \simeq l_D/3r_{\rm s}$, and thus we recover the well-known result that in the diffusion regime the cosmic-ray density scales as $r_{\rm s}^{-1}$.  

\begin{figure}[h!]
\centering
\includegraphics[scale=.9,angle=0]{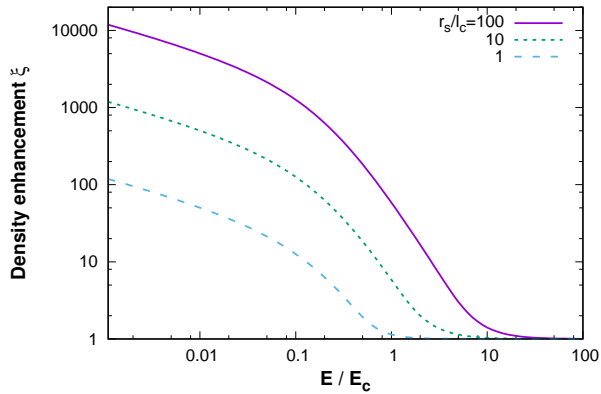}
\caption{Enhancement of the density of cosmic rays diffusing from sources at different distances.}
\label{fig:fluxenh}
\end{figure}

Another important consequence of the diffusion process is that the spectrum of the observed flux from a source at a given distance will be modified, with the lower-energy particles having their density enhanced by an amount that depends on the distance from the observer to the source and on the properties of the magnetic field. The  enhancement factor $\xi$ can be defined as the ratio between the actual density  and the one that would result in the case of rectilinear propagation, i.e. 
\begin{equation}
    \xi\equiv \frac{n(E,r_{\rm s})}{Q(E)/(4\pi r_{\rm s}^2c)}=\frac{1}{\langle \cos \theta \rangle}.
\end{equation}
This simple expression directly relates, for the case of a steady source, the dipolar-type anisotropy due to a source, which has an amplitude $\Delta=3 \langle \cos \theta \rangle$, to the diffusive enhancement of its flux, so that $\Delta=3/\xi$.

The enhancement factor  is shown in fig.~\ref{fig:fluxenh} for different values of the ratio of the source distance to the coherence length. This factor starts to be significantly larger than unity   for $E<E_{\rm rect}$, which happens at different values of $E/E_{\rm c}$ for sources having different values of $r_{\rm s}/l_{\rm c}$.  For example, for the farthest source considered in fig.~\ref{fig:fluxenh}, lying at $r_{\rm s} =100~l_{\rm c}$, the enhancement appears for $E \leq  10~E_{\rm c}$ and scales as $ (E/E_{\rm c})^{-2}$ for $E/E_{\rm c} \gg 0.2$ and as $(E/E_{\rm c})^{-1/3}$ for $E/E_{\rm c} \ll 0.1$, as expected from eq.~(\ref{D(E).eq}).  On the other hand, for the closest source considered, at $r_{\rm s}=l_{\rm c}$, the enhancement starts to be significant only for $E\leq E_{\rm c}$. Note that the enhancement factor $\xi$ can reach several orders of magnitude in some cases.

\subsection{Source emitting since redshift $z_i$}

  We have considered up to now the simple case of a stationary source, assuming an infinite time since the source started to accelerate the particles.  If the source were instead to start its emission at some time in the past, the number of particles reaching the observer from the source may be suppressed at low energies due to a magnetic horizon effect \cite{le04,be06,gl07,mo13}. This is due to the fact that, as the energy decreases, the time required for the diffusing particles to reach the observer  may become larger than the source lifetime. This effect could clearly be more relevant if the lifetime of the source is much smaller than the age of the Universe. In the previous discussion we also neglected the effects of the expansion of the Universe and of the energy losses suffered by the particles when they interact with the background radiation fields, which  affect the spectrum observed from a source  mostly at the highest energies  (for CRs with Lorentz factors $\Gamma>10^9$ in the case of interactions with the CMB).

In order to study the more general case, consider the density $n$ of ultrarelativistic particles propagating from a source located at ${\vec r}_s$ in an expanding Universe, which in the diffusion regime obeys the equation \cite{be06}

\begin{eqnarray}
\frac{\partial n}{\partial t}+3 H(t) n - b(E,t) \frac{\partial n}{\partial E} - n \frac{\partial b}{\partial E} - \frac{D(E,t)}{a^2(t)} \nabla^2 n 
 =\frac{Q(E,t)}{a^3(t)} \delta^3({\vec r}-{\vec r}_s),
\label{difeq}
\end{eqnarray}
where ${\vec r}$ denotes the comoving coordinates, $a(t)$ is the scale factor of the expanding Universe, $H(t) \equiv {\dot a}/a$ is the Hubble constant and $D(E,t)$ is the diffusion coefficient. The  source function $Q(E,t)$ gives the number of particles emitted per unit energy and time. The energy losses of the particles are described by 
\begin{equation}
\frac{{\rm d}E}{{\rm d}t}=-b(E,t),\ \ \ \ \ \ b(E,t)=H(t) E +b_{\rm int}(E).
\label{eloss}
\end{equation}
This includes the energy redshift due to the expansion of the Universe and energy losses due to the interactions with the radiation backgrounds, which in the case of protons include pair production and photopion production due to interactions with the CMB. For heavier nuclei the main interaction effect is their photodisintegration, both off the CMB and off the extragalactic background light (EBL), so that a generalization of this equation to include different coupled species needs in principle to be considered \cite{abg1,abg2}. The general solution for the case of protons was obtained by Berezinsky and Gazizov \cite{be06,be07}, being
\begin{equation}
n(E,r_{\rm s})=\int_0^{z_{i}}{\rm d}z\,\left|\frac{{\rm d}t}{{\rm d}z}\right| Q(E_g,z)\frac{\exp \left[-r_{\rm s}^2/4\lambda^2\right]}{(4\pi\lambda^2)^{3/2}} \frac{{\rm d}E_g}{{\rm d}E},
\label{siro.eq}
\end{equation}
where  $z_{i}$ is the initial redshift when the source started to emit (which in the diffusive regime has the meaning of time rather than distance) and $E_g(E,z)$ is the original energy at redshift $z$ of a particle having energy $E$ at present ($z=0$). The source function  $Q$ will be assumed for definiteness to correspond to a power-law spectrum, $Q \propto E_g^{-\gamma_s}$, and it may eventually have an overall evolution with redshift.  The Syrovatsky variable is given by
\begin{equation}
\lambda^2(E,z)=\int_0^{z}{\rm d}z'\,\left|\frac{{\rm d}t}{{\rm d}z'}\right| (1+z')^2\,D(E(z'),z'),
\nonumber
\end{equation}
with $\lambda(E,z)$ having the meaning of the typical distance over which the CRs diffuse from the site of their production with energy $E_g(E,z)$ at redshift $z$ until they are degraded down to energy $E$ at the present time. In the expanding Universe
\begin{equation}
\left|\frac{{\rm d}t}{{\rm d}z}\right|=\frac{1}{H_0(1+z)\sqrt{(1+z)^3\Omega_m+\Omega_\Lambda}},
\nonumber
\end{equation}
where we consider the present values $H_0\simeq 70$~km\,s$^{-1}$Mpc$^{-1}$ for the Hubble constant, $\Omega_m\simeq 0.3$ for the matter content and $\Omega_\Lambda\simeq 0.7$ for the cosmological constant contribution. 

The general effect of considering an initial redshift $z_i$ at which the particle acceleration started  will be to deplete the flux at low energies with respect to that expected from fig.~\ref{fig:fluxenh}. 
In order to study this we will consider the case in which $Q(E,z)$ vanishes at  redshifts higher than $z_i$ and remains constant after that time, focusing on a source emitting since relatively recent times, i.e. with $z_i<0.2$. 

\begin{figure}[h!]
\centering
\includegraphics[scale=.44,angle=-90]{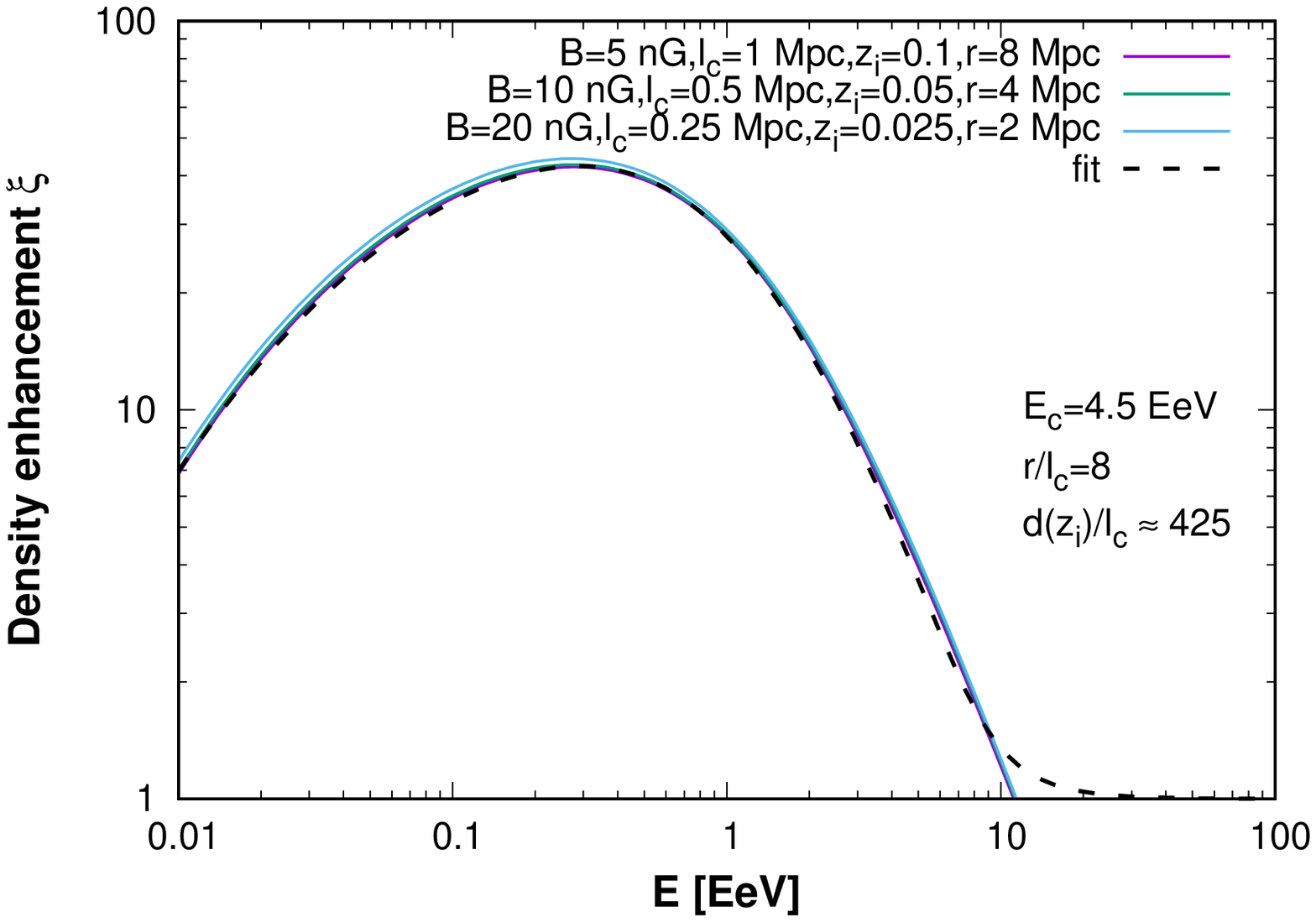}
\includegraphics[scale=.44,angle=-90]{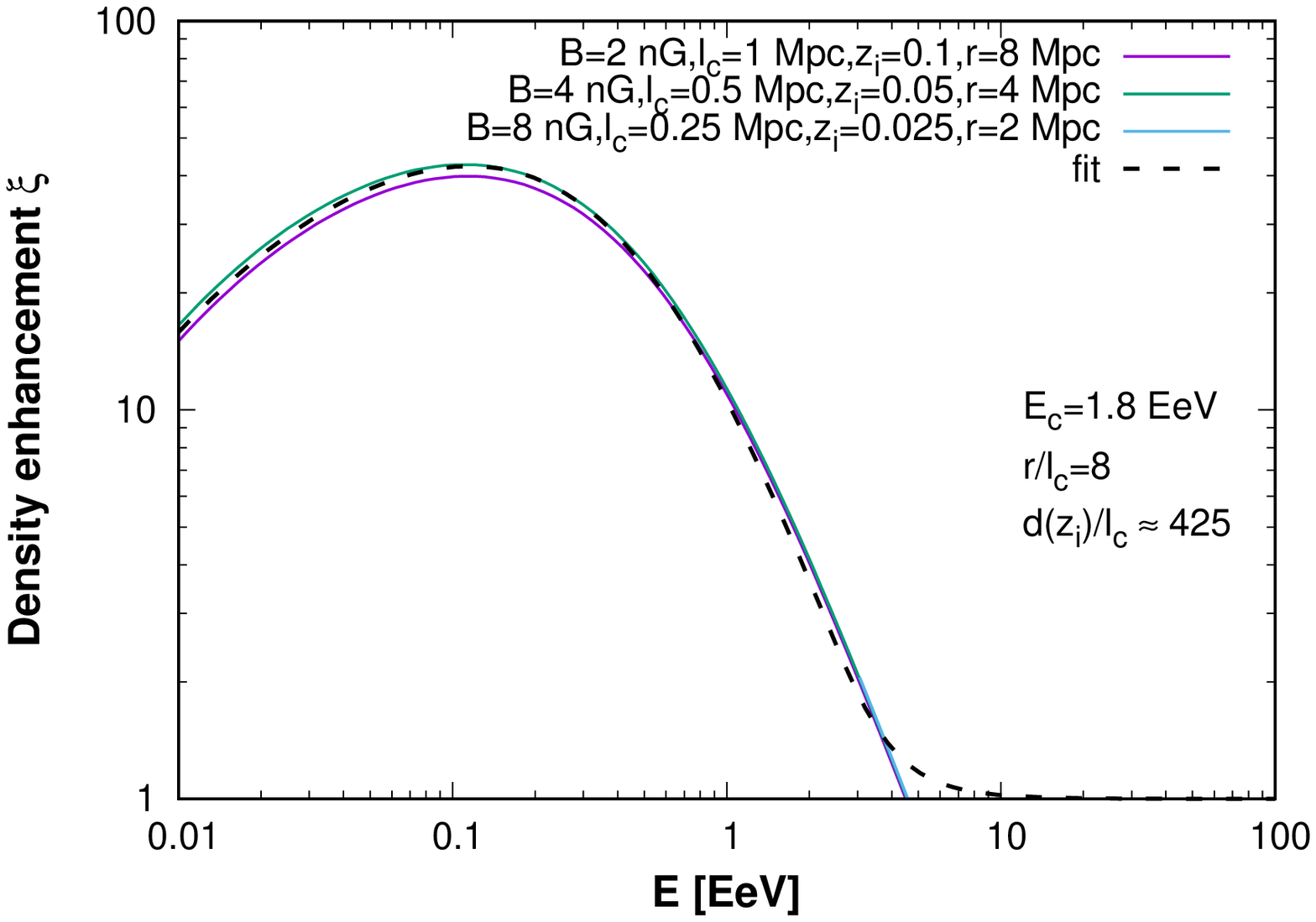}
\includegraphics[scale=.44,angle=-90]{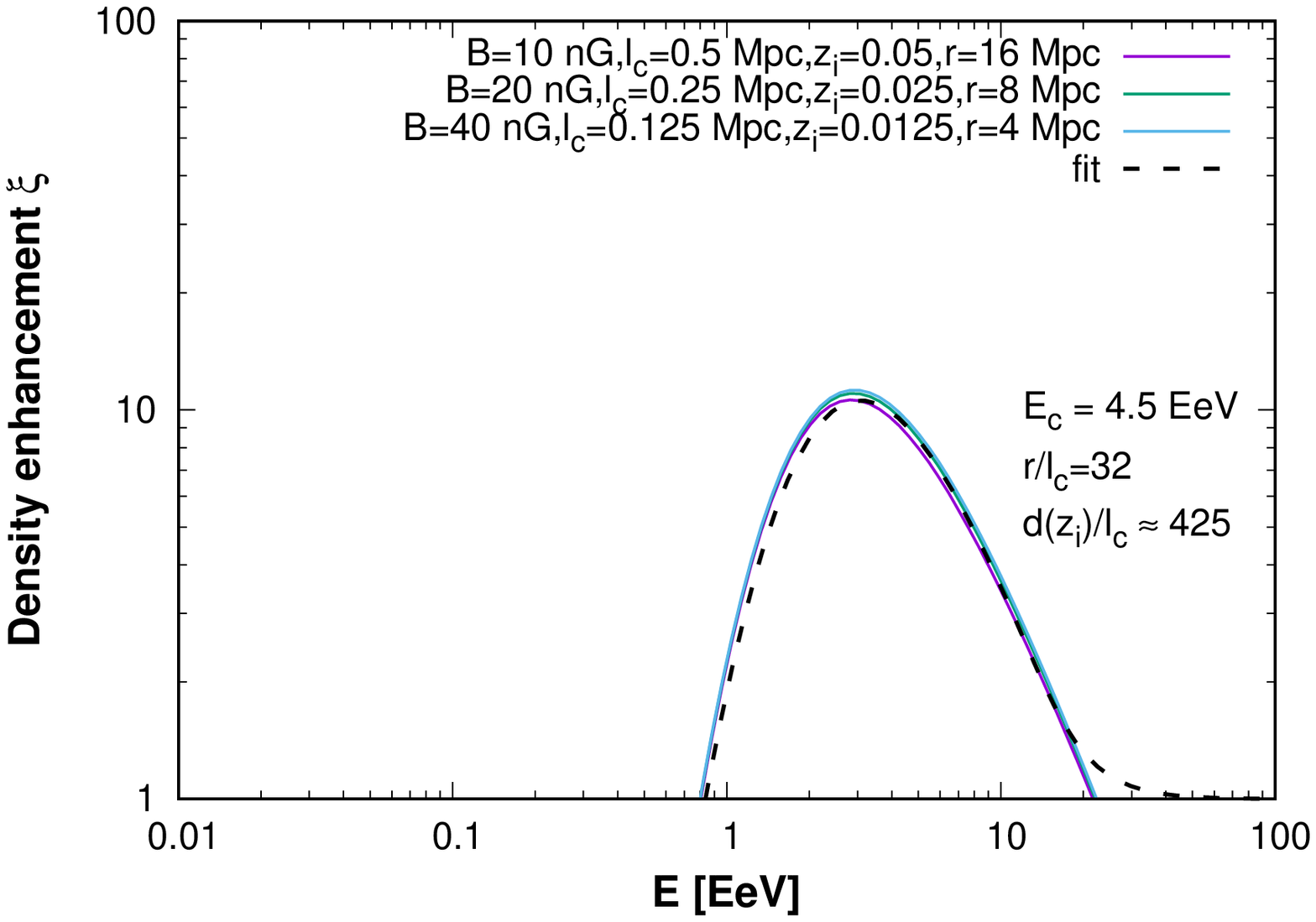}
\includegraphics[scale=.44,angle=-90]{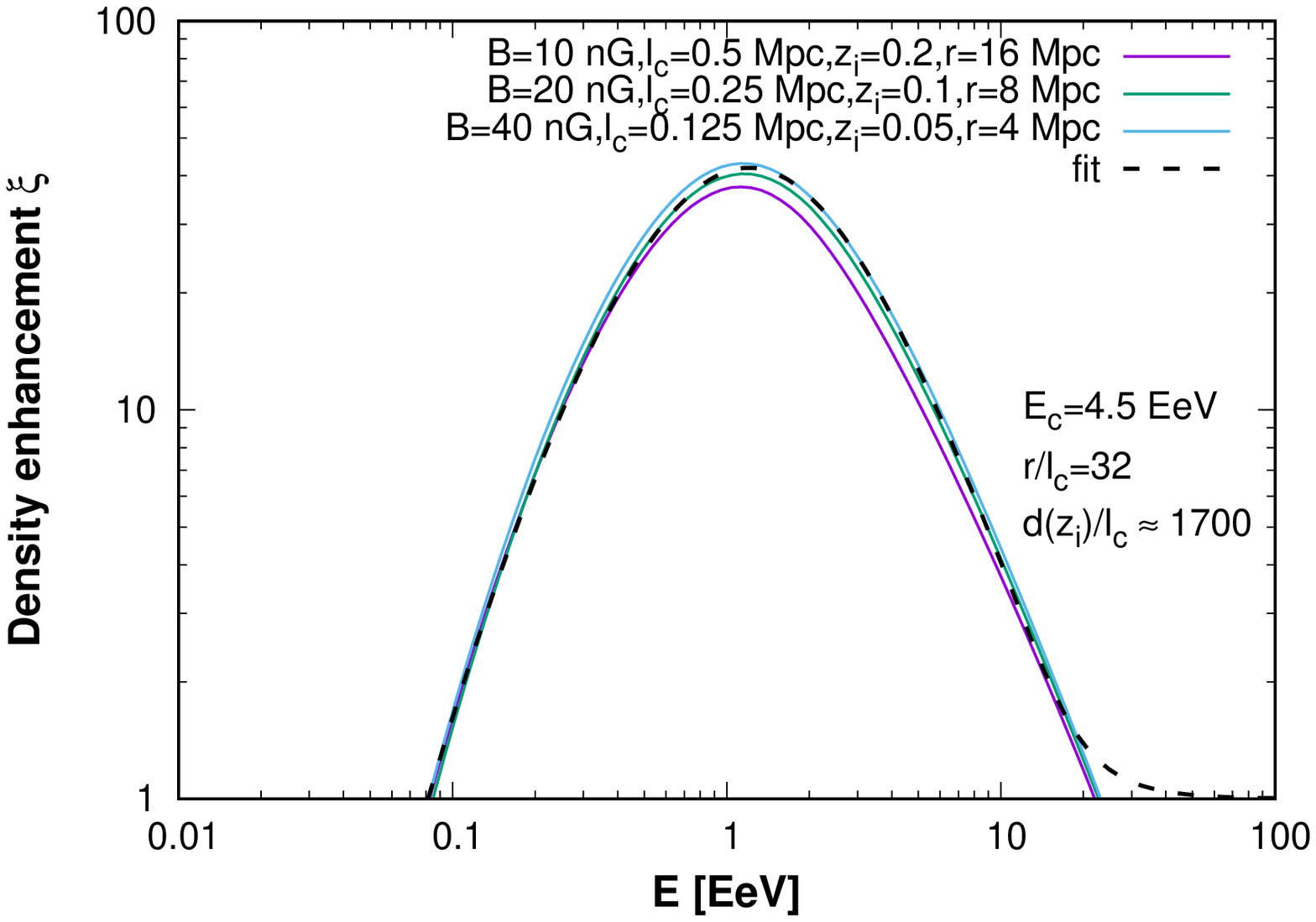}
\caption{Enhancement of the CR density for a proton source and for different values of the relevant parameters. Also shown are the fits obtained using eq.~(\ref{nerp}). We used in these plots $\gamma_s=2$, but the results are quite insensitive to the actual spectral index adopted.}
\label{fig:fluxenhp}
\end{figure}

The results for a source of protons and  for different values of the parameters, obtained by numerical integration of eq.~(\ref{siro.eq}), are shown in fig.~\ref{fig:fluxenhp}. The main parameters determining the low-energy suppression of the density enhancement factor  are the distance to the source $r_{\rm s}$, the maximum redshift $z_i$, the magnetic field amplitude $B$ and its coherence length $l_{\rm c}$\footnote{The coherence length is assumed to be stretched by the expansion, so that $l_{\rm c}(z)=l_{\rm c}(0)/(1+z)$, while MHD considerations suggest \cite{be07} that $B(z)=(1+z)B(0)$.}. 
It can be seen that there are some combinations of these parameters that lead to very similar results, with the relevant independent combinations  being $E_{\rm c}$, $r_{\rm s}/l_{\rm c}$ and $d(z_i)/l_{\rm c}$, where $d(z_i)$ is the total comoving distance traveled by the particles emitted at the initial  redshift $z_i$, 
\begin{equation}
    d(z_i)= \int_0^{z_i}{\rm d} z \frac{c}{H_0\sqrt{(1+z)^3\Omega_m+\Omega_\Lambda}} \simeq r_{\rm H}(z_i-0.225~ z_i^2),
\end{equation}
where $r_{\rm H}\equiv c/H_0\simeq 4.3$~Gpc is the Hubble radius.

The resulting density can be written in this case as
\begin{equation}
n(E,r_{\rm s}) = \frac{Q(E)}{4\pi c r_{\rm s}^2 }\xi_i,
\label{nerp}
\end{equation}
where an accurate fit to the enhancement factor is
\begin{equation}
    \xi_i\simeq \frac{1}{\ C(r_{\rm s}/l_D) } \exp \left[-\left(\frac{r_{\rm s}^2}{0.7 l_D(E) d(z_i)}\right)^{0.82}\right],
\label{xim}
\end{equation}
with the function $C(R)$  given by eq.~(\ref{cos}).

\begin{figure}[h!]
\centering
\includegraphics[scale=.4,angle=-90]{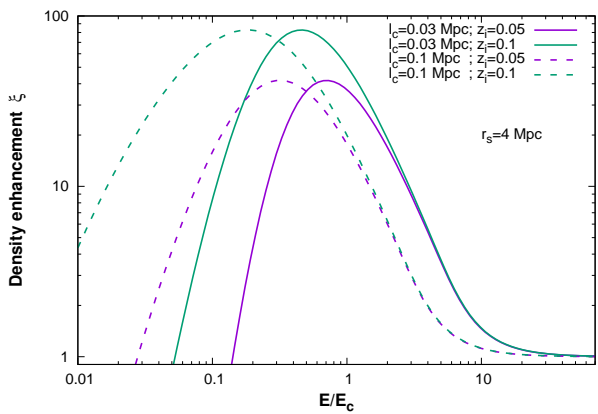}
\caption{Enhancement factor $\xi$ as a function of $E/E_{\rm c}$ for a fixed source distance of 4~Mpc, considering different values of $l_{\rm c}$ and $z_i$. .}
\label{fig:xsivsedec}
\end{figure}

The fitted enhancement factor
is plotted as a black dashed line in fig.~\ref{fig:fluxenhp} for each of the cases considered. It can be seen that the fit in eq.~(\ref{xim}) closely follows the solution of the diffusion equation in the energy range where diffusion holds, tending to the unmodified source spectrum  in the rectilinear propagation regime. 
For fixed $r_{\rm s}/l_{\rm c}$ and $d(z_i)/l_{\rm c}$ the results just depend on $E/E_{\rm c}$, while for fixed $E_{\rm c}$ the effect of increasing $r_{\rm s}/l_{\rm c}$ or decreasing $d(z_i)/l_{\rm c}$ is to produce a stronger suppression of the flux at low energies.
It is easy to see from eq.~(\ref{xim}) that the maximum of the flux enhancement happens at an energy such that $l_D(E_{\rm max}) \simeq 1.1 ~r_{\rm s}^2/d(z_i)$, and that the enhancement factor at that energy is $\xi_i^{\rm max} \simeq 0.8~ d(z_i)/r_{\rm s}$. To better illustrate these features we  display in  fig.~\ref{fig:xsivsedec} the enhancement factor $\xi_i$ as a function of $E/E_{\rm c}$ for a fixed source distance of 4~Mpc, considering different values of $l_{\rm c}$ and $z_i$. One can appreciate from the results that, for a fixed $r_{\rm s}$,  the height of the maximum enhancement is just proportional to $z_i$, the energy $E_{\rm max}$ is slightly below $E_{\rm c}$, scaling approximately as  $E_{\rm max}\simeq 0.5E_{\rm c} r_{\rm s}/\sqrt{d(z_i)l_{\rm c}}$, the energy of the transition towards the rectilinear regime is at $E_{\rm rect}\simeq E_{\rm c}\sqrt{r_{\rm s}/l_{\rm c}}$ and the shape of the enhancement curve becomes wider for increasing values of $l_{\rm c}$.

Note that as long as the enhancement of the proton flux due to the diffusion takes place for energies below about 20~EeV, which is the regime in which we will be interested in this work, and as is indeed the case in the examples shown in fig.~\ref{fig:fluxenhp},  the effects of the interactions turn out to be negligible  for the  relatively nearby sources considered.  Even though  at lower energies the distances traveled are, due to the diffusion,  larger than the distance to the source $r_{\rm s}$, as long as $z_i$  is not larger than about 0.2 they  will still be smaller than the corresponding interaction lengths, which are larger than about a Gpc at these energies.  In this case, also the cosmological evolution effects in the solution of the diffusion equation will not be large.

The inclusion of an initial  redshift at which the source started to emit particles also has an impact on the amplitude of the observed dipolar anisotropies in the arrival directions. In the case of a steady source, for which $\Delta = 3 \langle \cos \theta \rangle \simeq 3~ C(r_{\rm s}/l_D)$,  the more isotropic part of the flux comes from the particles that originated at the highest redshifts. Thus, an increment of the anisotropy with respect to this value should result at small energies when an initial redshift for the emission of the source is introduced. In the diffusive regime, the dipole amplitude can be computed from the general expression $\Delta = l_D |\nabla n|/n$, using the solution to the diffusion equation with initial redshift $z_i$  given in eq.~(\ref{siro.eq}). Considering directly the density in  eq.~(\ref{nerp}) with the approximate fit in eq.~(\ref{xim}),  the anisotropy due to a source emitting since a maximum redshift $z_i$ can be expressed as
\begin{equation}
    \Delta \simeq 3~ C(r_{\rm s}/l_D) \left[1+1.64 \left(\frac{r_{\rm s}^2}{0.7 l_D d(z_i)}\right)^{0.82}\right],
    \label{dipzm}
\end{equation}
which is quite accurate as long as $d(z_i)>r_{\rm s}$.

\begin{figure}[h!]
\centering
\includegraphics[scale=.8,angle=0]{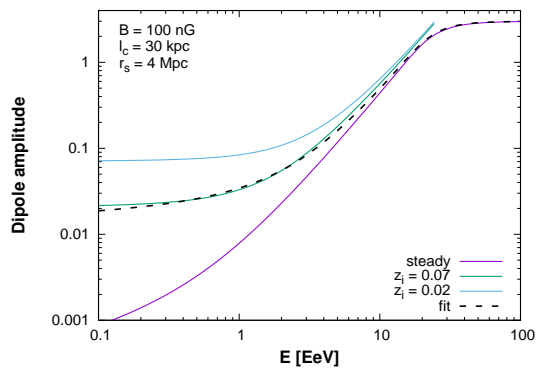}
\caption{Dipole amplitude as a function of the energy for a source at 4~Mpc
and a turbulent magnetic field of 100~nG amplitude and 30~kpc coherence length. The cases of a steady source and those with initial emission redshifts of 0.07 and 0.02 are  displayed, together with the fit from eq.~(\ref{dipzm}) to the $z_i=0.07$ case.}
\label{fig:dipoleamp}
\end{figure}

We show in fig. \ref{fig:dipoleamp} the dipole amplitude for the case of a steady source and also for two values of the initial redshift of emission, as well as the corresponding amplitude from eq.~(\ref{dipzm}) for the case with $z_i=0.07$. Let us note that the amplitude of the dipole at the energy for which the flux enhancement takes its maximum value is $\Delta(E_{\rm max}) \simeq 3.4 ~r_{\rm s}/d(z_i) \simeq 2.7/\xi_i^{\rm max}$.

The case of heavier nuclei can be described in a  similar way, as long as the interactions with the radiation backgrounds can be neglected, by just replacing the proton energy $E$ in the above expressions by the ratio $E/Z$, which is proportional to the rigidity of the particles. We will adopt this approximation in the following, but one should keep in mind  that the attenuation length for nuclear photodisintegration becomes smaller than 10~Mpc for energies larger than about  40~EeV for He or about 200~EeV for Fe nuclei. Hence, even for sources closer than 10~Mpc the attenuation could start to become non-negligible above those energies  in the regime of quasirectilinear propagation (or actually also  for somewhat lower energies if the CRs are still diffusing).

\subsection{Bursting source scenario}

Another potentially interesting scenario could be one in which the nearby source had a  burst of activity in the past, at redshifts between  $z_i$ and  $z_f$, but remained inactive afterwards \cite{be90,wa96}. 
In this case, the CRs reaching the observer would have traveled a total distance of at least $d(z_f)\simeq r_{\rm H}z_f$ (and at most $d(z_i)\simeq r_{\rm H}z_i$). The results will differ from those discussed in  the previous subsection  if $d(z_f)\gg r_{\rm s}$ since, in this case, the high-energy particles that are in the quasirectilinear regime would have already passed by the observer at the time of observation. This should lead to a stronger suppression of the spectrum at the highest energies. On the other hand, the absence of the high-energy particles that could have arrived through straighter trajectories, and hence more anisotropically,  implies that the dipolar amplitude should decrease when $d(z_f)\gg r_{\rm s}$, especially at high energies (while the effect of a finite $z_i$ was instead to enhance the anisotropies at low energies, as shown in fig.~\ref{fig:dipoleamp}).  

\begin{figure}[h!]
\centering
\includegraphics[scale=.88]{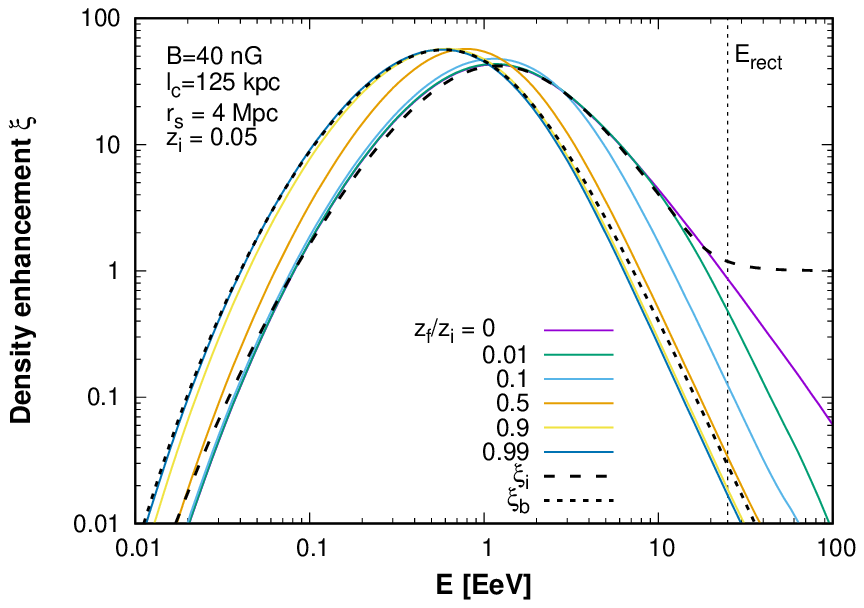}
\includegraphics[scale=.88]{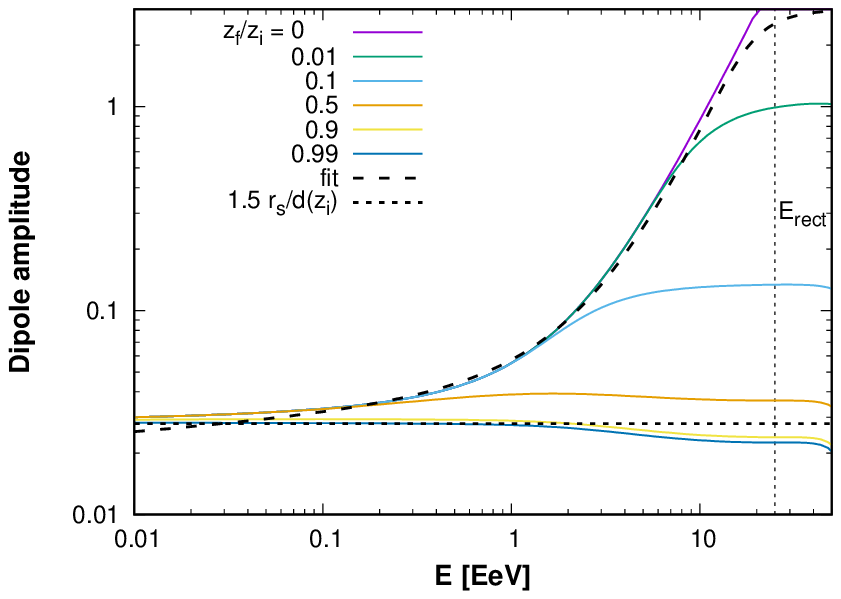}
\caption{Left panel: Enhancement of the CR density for a proton source and for different burst durations (the  values of the relevant source and magnetic field parameters are also indicated). Right panel: Associated dipole amplitudes for different burst durations. A vertical line indicates the reference energy $E_{\rm rect}$.}
\label{fig:burst}
\end{figure}

We show in the left panel of fig.~\ref{fig:burst} the density enhancement factors for the bursting sources, considering for illustration the values $B=40$~nG, $l_c=125$~kpc and $r_s=4$~Mpc,  adopting $z_i=0.05$ and for different burst durations with   $z_f/ z_i=0.01$, 0.1, 0.5, 0.9 and 0.99. Since for bursting sources the CR density does not tend  towards the steady rectilinear case at the highest energies, the normalization of the enhancement factors turns out to be somewhat arbitrary. We normalize them here  such that the total number of emitted particles  is similar in all cases, i.e., we adopt $Q(E)\,\Delta T=N(E)$, with the burst duration being $\Delta T=[d(z_i)-d(z_f)]/c$. One can see that  as $z_f$ gets closer to $z_i$  the maximum of the curves get shifted to lower energies, by up to a factor of about 2 for the shortest burst, and the spectrum gets increasingly steeper at  higher energies. On the other hand, at energies below the maximum the spectral suppression related to the magnetic horizon effects, due to the absence of emission at $z>z_i$, keeps a similar shape in all cases. 
Let us mention that the spectrum would also be essentially cut off at energies such that the minimum distance traveled by the CRs from the source, $d(z_f)$, becomes larger than the attenuation length associated with the propagation through the photon backgrounds. 

In the right panel of  fig.~\ref{fig:burst}, we show the amplitudes of the associated dipolar type anisotropies  in these scenarios, with the main change being the flattening of the amplitudes that appear at the highest energies as the burst duration is reduced. 

An interesting limiting case is that of a source having an `instantaneous' burst at a redshift $z_b$, i.e. with $Q(E,z)\propto \delta(z-z_b)$. In this  case, if one neglects the effects of the interactions and of the cosmological evolution,  the  enhancement factor is given by
\begin{equation}
    \xi_b\propto \frac{\exp\left[-3r_{\rm s}^2/(4l_D\,d(z_b))\right]}{[l_D\,d(z_b)]^{3/2}}.
    \label{xib}
\end{equation}
This enhancement reaches a maximum for $l_D=r_{\rm s}^2/2d(z_b)$.
The dipole amplitude is given by $\Delta=l_D|\nabla n|/n=l_D ({\rm d\,ln}\xi_b/{\rm d}r_{\rm s})=1.5~ r_{\rm s}/d(z_b)$, and hence it turns out to be energy independent. This value is indicated as a horizontal dotted line in the right panel of  fig.~\ref{fig:burst}, with the difference with respect to the results of the very short burst with $z_f=0.99z_i$ being due to the attenuation effects that were included in this last case. This indicates that for a scenario with one dominant source which had a short burst in the past, in order to get anisotropies below $\sim 10$\% (as is indeed observed at 10~EeV), the CRs from the source should have traveled a distance larger than about $15r_{\rm s}$ (corresponding to $z_b>15 r_{\rm s}/r_{\rm H}$). This requirement could be in tension with the fact that CRs with energies above 150~EeV have been observed since, no matter which composition is assumed, CRs could not have arrived with those energies after travelling more than 50~Mpc through the background radiation. Hence, in scenarios with just a short burst  of emission and in which the CRs above few EeV and up to the highest energies originate from the nearby source, the source would need to be not much farther than $\sim 4$~Mpc.

The results discussed in this section could be relevant for the interpretation of the observations of the spectrum and the composition of the cosmic rays at ultrahigh energies. Indeed, taking into account the enhancement of the intergalactic magnetic fields in the Local Supercluster, a powerful source in our neighborhood may give rise to a significant contribution to the cosmic-ray fluxes 
  above few EeV. In particular,  the spectral hardening observed at the ankle may turn out to be the result of the light components of the nearby source, i.e. the H and He, getting suppressed just below the ankle as  a result of the energy dependence of the diffusion effects, while the heavier components could be emerging at higher energies as their magnetic horizon ceases to be effective. The hardening of the spectra of the different mass components at low rigidities may be directly related to the maximum redshift at which the nearby source started to emit significant amounts of UHECRs, and it would depend on  the actual distance to the source since the suppression becomes strong when the condition $l_D\ll r_{\rm s}^2/d(z_i)$ is satisfied.
In the following,  we describe scenarios of this kind and compare their predictions to the observed cosmic-ray spectrum and composition. Moreover, we also study the contribution of the nearby source to the UHECR dipolar-type anisotropies, showing that it can naturally account for the measured dipole amplitudes at different energies.

\section{A scenario with a strong nearby extragalactic source}

\subsection{The flux from the nearby source}
We consider here scenarios in which a nearby extragalactic UHECR source provides a significant fraction of the cosmic-ray flux observed at the Earth above EeV energies, as a consequence of the large density enhancement resulting from the diffusion of its cosmic rays in the strong Local Supercluster magnetic field. We model this field as having a spatially homogeneous root-mean-square strength and a Kolmogorov spectrum of turbulence, using the results of the previous sections\footnote{When estimating the flux from the nearby source, we do not consider the effects associated with the Local Supercluster being of finite size, which is a good approximation as long as the source is not close to the boundary of this region.}. 
We will model the flux from the nearby source adopting five representative mass components, $i={\rm H}$, He, N, Si and Fe. Their flux will then be parametrized as
\begin{equation}
\Phi^s(E,r_{\rm s}) = \Phi^s_0  \sum_i f^s_i\left(\frac{E}{\rm EeV}\right)^{-\gamma_s} \xi(E/Z_i,r_{\rm s})\frac{1}{\cosh(E/Z_iE^s_{\rm cut})},
\label{phis}
\end{equation}
where $\xi$ is the flux enhancement factor for the scenario adopted. We will consider the case with continuous emission since an initial redshift $z_i$ (i.e. with $\xi=\xi_i$ from eq.~(\ref{xim})) and  also the one with a very short burst of emission at a redshift $z_b$ (with $\xi=\xi_b$ from eq.~(\ref{xib})).
The overall strength of this source, $\Phi^s_0$, as well as its spectral slope, $\gamma_s$, and the relative fractions of the different nuclei, $f^s_i$, will be determined so as to account for the experimental  measurements. We also allow for an eventual rigidity-dependent energy at which the acceleration at the sources is cut off, leading to an effective exponential suppression of the fluxes observed at the Earth above energies $Z_iE^s_{\rm cut}$ (the cosh function allows to smoothly match the power-law at lower energies with the exponential suppression).

\subsection{Modelling the diffuse extragalactic contribution}

Besides the local source, there will certainly be a contribution from all the remaining UHECR sources in the Universe, which for simplicity will be considered to have similar spectral slopes and to be steady, adding up to a diffuse flux which will be assumed to be isotropic. In this case, since sources from very far away are expected to contribute sizably, their spectra will be significantly affected by the interactions with the radiation backgrounds. A practical way to take these effects into account is by introducing a modification factor $\eta$, defined as the ratio between the spectrum from the sources including the attenuation effects with respect to the spectrum that would have been expected from the same sources in the absence of interactions \cite{dip}. 
\begin{figure}[t]
\centering
\includegraphics[scale=1.3,angle=0]{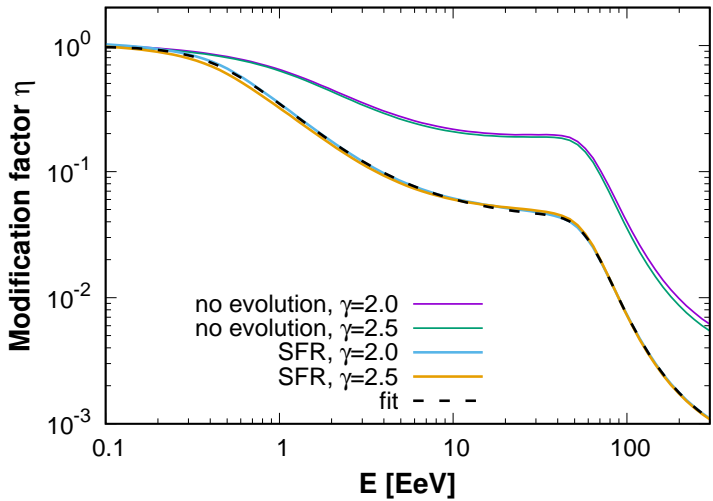}
\caption{Modification factor of the proton spectrum accounting for the interactions with the CMB photons, for two values of the spectral index and under two hypotheses for the evolution with redshift of the luminosity of the sources.}
\label{fig:eta}
\end{figure}

In the case of protons, the attenuation factor has been found to be quite insensitive to the source spectral index considered \cite{dip}, although it depends on the cosmological evolution adopted for the sources. In particular, we show in  fig.~\ref{fig:eta} the attenuation factors for spectra with $\gamma=2$ and 2.5 for the cases in which the sources do not evolve with redshift, i.e. for constant $Q(E)$, or for the case in which the sources evolve as the star formation rate (SFR).  For this last case, we consider the parametrization from \cite{ho06}, assuming that the source intensity evolves as $(1+z)^{3.44}$ up to redshift 0.97, evolving then as $(1+z)^{-0.26}$ for larger redshifts to then fall as $(1+z)^{-7.8}$ for redshifts above 4.48. The results in the plots actually include sources up to a maximum redshift of four, since the contribution from sources farther away is negligible. One can appreciate that, in comparison  with the no-evolution case, the main effect of the much larger number of sources that are present at high redshifts  in the case of the SFR evolution is to lead to a relative enhancement of the fluxes at energies below few~EeV (reflected in the fact that the values of $\eta$ are smaller by a factor of about 5 at higher energies). The computations are performed as in \cite{hmr14}.  One could mention also that the cosmological evolution of the SFR that we consider is somewhat  intermediate between that of active galaxies and that of gamma-ray bursts, and we will adopt it as a reference in the following since it can be expected to be more realistic than the scenario with no evolution.

We also include in fig.~\ref{fig:eta} an analytic fit to the attenuation factor obtained for the case of the star formation rate evolution, which accurately reproduces the results obtained. This fit can be used to model in an easy way the extragalactic diffuse  proton contribution by just convoluting the original power-law spectrum with the attenuation factor  (eventually adding a high-energy cutoff). The fitting function considered is
\begin{equation}
    \eta^{\rm H}(E)=\left[1+ 1/g_1(E)+1/g_2(E)\right]^{-1},
\end{equation}
where $g_1$ accounts for the effects of the photopion production interactions while $g_2$ for those of pair production (both with the CMB). They are parametrized  in terms of the functions 
\begin{equation}
    F_{[A,B,C]}(E)\equiv A\exp(B\,E^C),
\end{equation} 
with $E$ in EeV,  as $g_1(E)=F_{[0.0016,700,-1.2]}(E)$ and $g_2(E)=F_{[0.04,4.2,-0.46]}(E)+F_{[0.003,1.6,0.14]}(E)$.
The diffuse flux of hydrogen nuclei  will then be modelled as
\begin{equation}
    \Phi^H(E)=f_{\rm H}\Phi_0 \left(\frac{E}{\rm EeV}\right)^{-\gamma}\frac{\eta^{\rm H}(E)}{\cosh(E/E_{\rm cut})},
\end{equation}
where $f_{\rm H}$ is the hydrogen fraction in the diffuse flux at the low energies for which the modification factors are unity. We also allow for an effective exponential cutoff above the energy $E_{\rm cut}$. 

\begin{figure}[h!]
\centering
\includegraphics[scale=.8,angle=0]{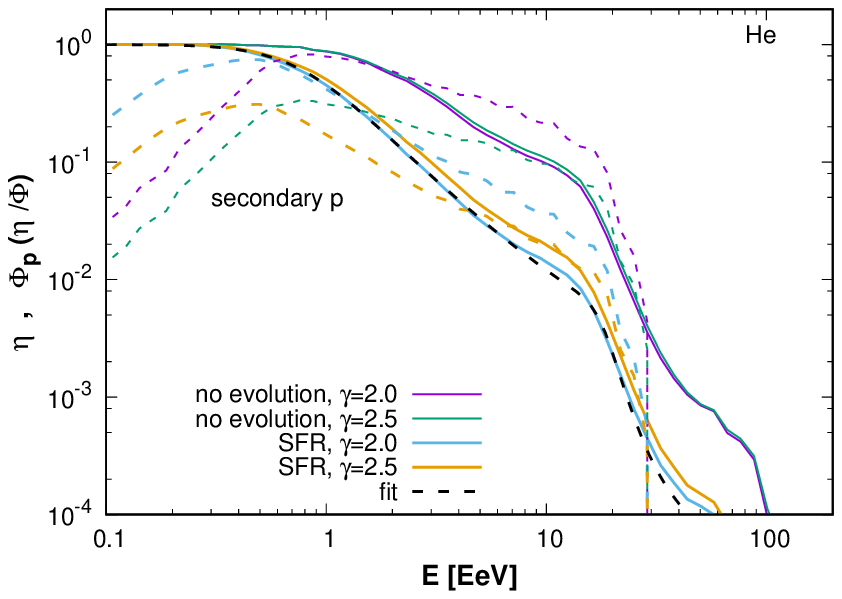}
\includegraphics[scale=.8,angle=0]{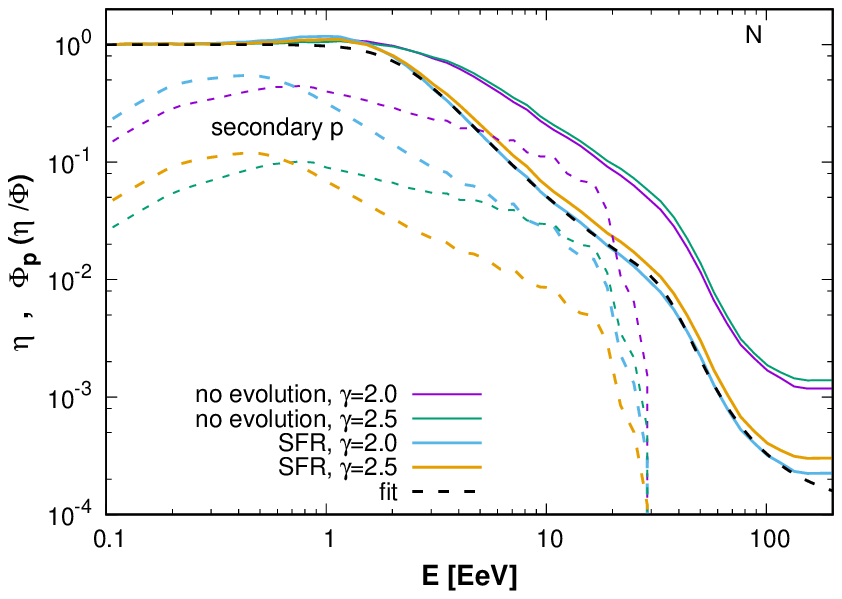}
\includegraphics[scale=.8,angle=0]{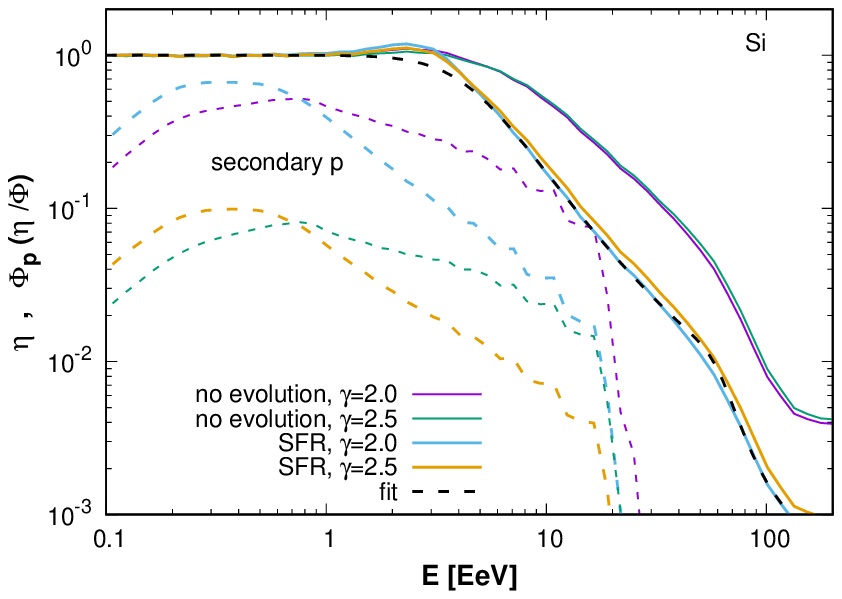}
\includegraphics[scale=.8,angle=0]{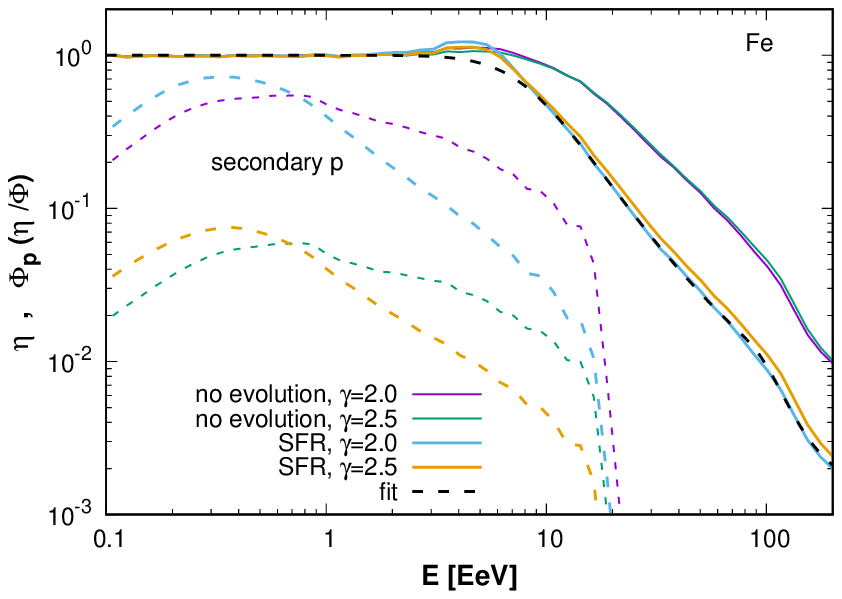}
\caption{Modification factor of the spectra of different  nuclei  (solid lines), accounting for the interactions with the photon backgrounds for two values of the spectral index of the source and two  hypothesis for the source luminosity evolution. The black dashed line corresponds to the fit to the modification factor, according to eq.~(\ref{eq:etaj}), for the evolution following the star formation rate and for $\gamma =2.$  Shown with dashed lines is the relative flux of secondary protons, $\Phi_p$ (note that the normalization factor $\Phi/\eta$ is just the flux of the primary nuclei that would be expected in the absence of interactions). }
\label{fig:etaheavy}
\end{figure}

Regarding the heavier nuclei, we will also consider just four mass groups, labelled as He, N, Si and Fe. We will assume that the sources inject these elements, as well as the hydrogen,  in  proportions characterized by the fractions $f_i$ that they contribute at a given energy (satisfying $f_{\rm H}+f_{\rm He}+f_{\rm N}+f_{\rm Si}+f_{\rm Fe}=1$). We will consider  that the acceleration depends on rigidity so that all species have the same spectral index at the sources and a common rigidity cutoff giving rise to an effective exponential suppression of the observed fluxes above energies $ZE_{\rm cut}$. 
The nuclei will be affected by their photodisintegrations off the photon backgrounds (which reduces the mass of the leading fragment and leads to the emission of secondary nucleons), as well as by electron-positron pair production (which reduces their Lorentz factor without changing their mass). Photopion production of heavy nuclei is only relevant for Lorentz factors larger than $4\times 10^{10}$, which corresponds to energies larger than those being considered here. We   collect all of the leading fragments heavier than H that result from the photodisintegration of a given primary element  in the mass group of that element, while the secondary protons are considered separately (the emitted neutrons will quickly decay into protons).  In this way, it is possible to introduce an effective attenuation factor for each mass group.
Note that some of the leading fragments from heavy nuclei may be light, but anyway the resulting mass distribution of the leading fragments is generally peaked close to the mass of the primary. The total spectrum can then be obtained by adding up the contributions from the different mass groups as well as the secondary protons.
On the other hand, when computing the average logarithmic mass and its dispersion we will use  the actual mass distribution of the leading fragments obtained in the simulations, since neglecting the spread in each mass group could lead to slight differences in the results. For these computations we follow \cite{hmrheavy}, using the photodisintegration cross sections from \cite{psb,salomon} and the redshift evolution of the EBL background from \cite{in13}. The attenuation factors obtained for different source spectral indices and redshift evolutions are shown in fig.~\ref{fig:etaheavy}. One can note that also in the case of nuclei the attenuation factors are quite insensitive to the value of the spectral index and that they do depend on the assumed source evolution.

For the reference source evolution following the star formation rate,  we also display in fig.~\ref{fig:etaheavy} an analytic fit to the suppression factor. 
The fits for the mass groups $j={\rm He}$, N, Si and Fe are performed with the functions
\begin{equation}
    \eta^j(E)=\left[1+ 1/g^j_1(E)+1/g^j_2(E)\right]^{-1},
    \label{eq:etaj}
\end{equation}
where now the different functions are just $g^j_i(E)=F_{[A_i^j,B_i^j,C_i^j]}(E)$. The functions $g_1^j$ account mostly for the effects of the photodisintegrations off the CMB while $g_2^j$ for those of the photodisintegrations  with the EBL, although the subdominant pair production effects are also included in them.
The different parameters of these fits are collected in Table~\ref{tab:fitnuclei}.

The total diffuse flux  from the five mass groups will then be 
\begin{equation}
\Phi^{\rm dif}(E) = \Phi_0  \sum_i f_i\left(\frac{E}{\rm EeV}\right)^{-\gamma} \eta^i(E)\frac{1}{\cosh(E/Z_iE_{\rm cut})},
\label{phid}
\end{equation}
where the sum runs over $i={\rm H}$, He, N, Si and Fe.

\begin{table}[ht!]
\centering
\caption{Coefficients of the fits to the attenuation factors for the different nuclei obtained by adopting a source luminosity evolution proportional to the star formation rate.}
\bigskip
\begin{tabular}{c c c c c c c }
\hline\hline
    Element & $A^j_1$ & $B^j_1$ & $C^j_1$ & $A^j_2$& $B^j_2$ & $C^j_2$ \\
 \hline
 He & $4.1\times 10^{-5}$ & $2.0\times 10^{3}$&-2.0 &$3.8\times 10^{-5}$ & 10 & -0.24\\
 N &$1.1\times 10^{-4}$  & $1.2\times 10^{3}$ &-1.5 &$6.8\times 10^{-4}$ &11 &-0.40 \\
 Si &$6.9\times 10^{-4}$  & $1.3\times 10^{5}$ &-2.5 & $3.4\times 10^{-4}$ & 14 & -0.34 \\
 Fe & $2.2\times 10^{-3}$ &$1.7\times 10^{6}$ & -2.8 & $3.4\times 10^{-3}$& 18 & -0.36\\
\end{tabular}
\label{tab:fitnuclei}
\end{table}

\begin{figure}[h!]
\centering
\includegraphics[scale=1.1,angle=0]{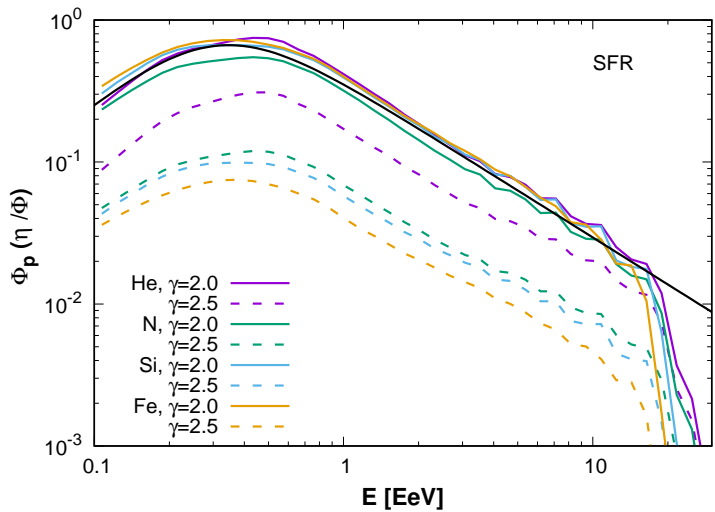}
\caption{Ratio of the spectrum of secondary protons to that of the primary nuclei for the star formation rate evolution case. Solid lines correspond to a spectral index $\gamma=2$ and dashed lines to $\gamma=2.5$.  The black solid line correspond to the fitting function $G(E)$ in  eq.~(\ref{ge}).}
\label{fig:secp}
\end{figure}

Regarding the secondary protons, one can see that they get produced in significant amounts (comparable in some cases to the primary fluxes) in the energy range  between  0.1 and few EeV. Their flux depends on the source spectral index and on the cosmological source evolution considered. Their maximum energies  are actually directly related to the maximum energies of the primaries as $E_{\rm max}^p=E_{\rm max}^j/A\simeq E_{\rm cut}/2$ (in these simulations we just consider for definiteness the case of a sharp cutoff at $E/A=30$~EeV). We also take into account that after the secondaries get produced and until they arrive at the Earth the proton energies get degraded, mostly due to pair production and to adiabatic redshift losses. For the reference source evolution case,  following the SFR, we collect the results of the secondary proton fluxes in fig.~\ref{fig:secp}. The density of secondary protons can be approximately fitted as
\begin{equation}
    \Phi^p(E)\simeq \Phi_0\sum_j f_j \left(\frac{E}{\rm EeV}\right)^{-\gamma}\frac{A^{2-\gamma}G(E)}{\cosh(2E/E_{\rm cut})},
    \label{secflux}
\end{equation}
where for the SFR evolution case one has
\begin{equation}
G(E)\simeq \frac{1}{2.7 (E/{\rm EeV})^{1.1}+0.15/(E/{\rm EeV})^{1.4}}.
\label{ge}
\end{equation}

The factor $A^{2-\gamma}$ in eq.~(\ref{secflux}) can be understood by noting that when there is a total disintegration of a nucleus with energy $E$, this produces $A$ nucleons with energy $E/A$. Hence, the amount of secondaries at a given energy depends, in a first approximation, on the flux of primaries in a similar logarithmic energy bin at an energy $A$ times larger, so that their relative ratio at a given energy should scale as $A \times A^{-\gamma+1}$. The fact that the disintegration may not be complete and that the protons suffer energy losses as they propagate makes this relation not exact, although it is still quite accurate.

Finally, we also include a Galactic contribution, consisting mostly of heavy elements. This component is relevant only below few EeV but is anyhow subdominant above 0.1~EeV. It is taken from \cite{mr19}, considering the results that include an exponential Galactic cutoff at energies $\sim 40Z$~PeV.

\section{Results}
In this section we present the expectations from two illustrative scenarios, one with a source emitting continuously since an initial redshift $z_i$ and another having a short burst of emission at redshift $z_b$. Both of them   account for the main observations of the spectrum and composition obtained by the Auger Collaboration above $\sim 0.3$~EeV \cite{augersp17,augerxm}. Data from other experiments in this energy range also exist, relying on a smaller number of events  and having  different systematic effects, but we do not attempt to do a global fit to all of them. These scenarios  also account for the large-scale dipolar anisotropies observed above 4~EeV \cite{dipoleapj}. 

For definiteness,  for these examples we  adopt a source distance $r_{\rm s}=4$~Mpc,  similar to that of potential source candidates such as Cen~A or some starburst galaxies like NGC~4945 or M82.

\begin{figure}[t!]
\centering
\includegraphics[scale=1.3,angle=0]{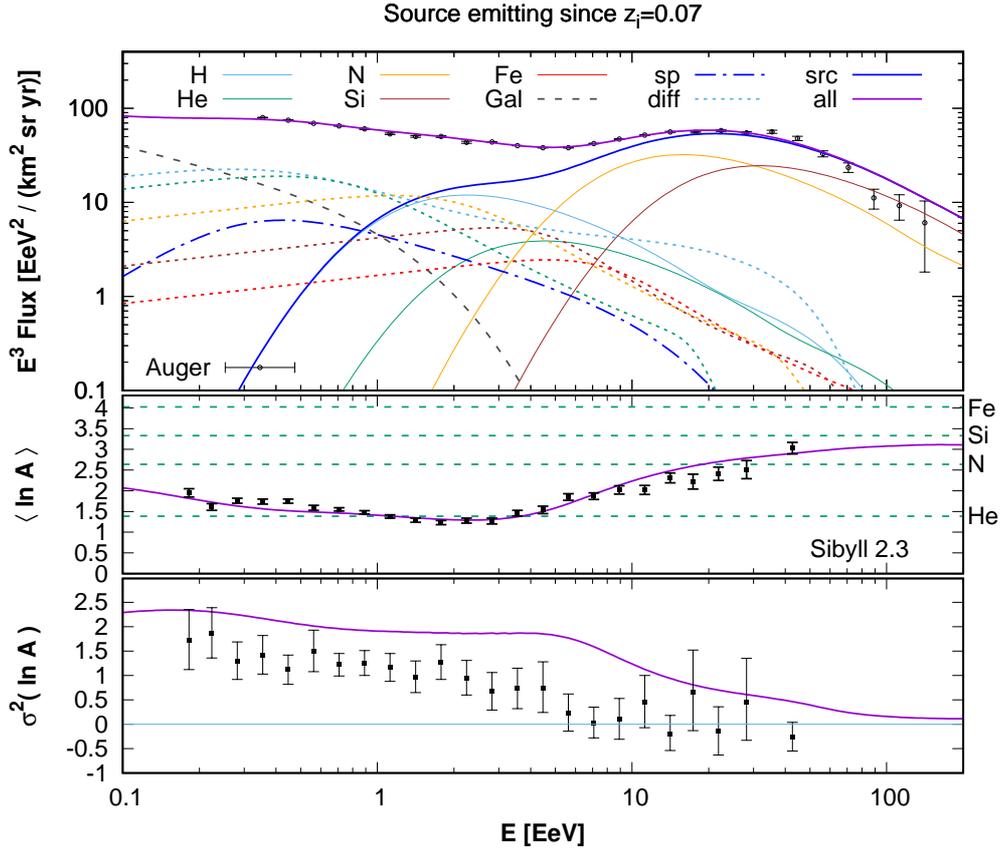}
\caption{ Energy spectrum (top panel), average logarithmic mass (middle panel) and variance of ln$A$ (bottom panel) vs. energy for the model, compared with data from the Pierre Auger Observatory \cite{augersp17,augerxm}. The different spectral components correspond to the five mass groups from the nearby source (solid lines), those from the diffuse flux (dotted lines), the diffuse secondary protons (dot-dashed line) and the total Galactic contribution (dashed line). Also shown are the total  contribution from the nearby extragalactic source (src) as well as the sum of all components (all).
The experimental data on $\langle{\rm ln}A\rangle$ and $\sigma^2({\rm ln}A)$ are those obtained adopting the Sibyll~2.3 hadronic model.}
\label{fig:spcomp}
\end{figure}

In fig.~\ref{fig:spcomp} we show the expectations from the example with a source emitting  continuously  since an initial redshift $z_i$. We adopt $z_i=0.07$, a value leading to a maximum enhancement for the nearby source $\xi_{\rm max}\simeq 60$, which ensures that  the modulation effects due to the diffusion and magnetic horizon will be sizable and that the anisotropies will be sufficiently low.  This maximum redshift is also comparable to the estimates of the merging time of Cen~A that took place several hundred million years ago \cite{is98}. We will adopt a turbulent magnetic field in the Local Supercluster with  strength $B=100$~nG and a coherence length  $l_{\rm c}=0.03$~Mpc.\footnote{Note that the small value adopted for $l_{\rm c}$ is required in order that the shape of the enhancement factor be narrow (see fig.~\ref{fig:xsivsedec}), since this allows to reduce the overlap between different mass components and hence provides a better fit to the data.
} This leads to $E_{\rm c}\simeq 2.7$~EeV,  so that the light components H and He will be strongly enhanced  below the ankle and the heavier ones are enhanced above it. (Actually, in this example the enhancement of the different  components peaks at energies of about $2Z$~EeV).\footnote{The results would however remain similar if one were to rescale $r_{\rm s}$, $l_{\rm c}$ and $z_i$ by a factor $\lambda$ and divide $B$ by the same factor. In particular, for a source in the Virgo cluster at a distance of about 16~Mpc, similar results would be obtained for $B=25$~nG, $l_{\rm c}=0.12$~Mpc and $z_i=0.28$.} 
 The source parameters considered in this figure, which are chosen so as to approximately reproduce the experimental results,  are a spectral slope for the nearby source of $\gamma_s=2.3$, with relative fractions $f^s_{\rm H}=0.45$, $f^s_{\rm He}=0.09$, $f^s_{\rm N}=0.31$, $f^s_{\rm Si}=0.15$ and $f^s_{\rm Fe}=0$. For the diffuse flux we adopt   $\gamma=2.7$, with relative fractions $f_{\rm H}=0.45$, $f_{\rm He}=0.33$, $f_{\rm N}=0.15$, $f_{\rm Si}=0.05$ and $f_{\rm Fe}=0.02$, which allow to approximately reproduce the observations below 1~EeV.
We consider  upper rigidity  cutoffs with $E^s_{\rm cut}=E_{\rm cut}=20$~EeV, although the results do not depend significantly on the values adopted as long as these cutoff energies are larger than 5~EeV.

One can see that the model considered reproduces the main features of all the observations. The nearby source contributes significantly to the flux above 1~EeV, becoming actually dominant above about 5~EeV. The H and He components are strongly enhanced by the diffusion up to about the ankle energy of $\sim 5$~EeV. At higher energies, the heavier components become increasingly dominant and the final suppression above $\sim 40$~EeV is related to the energy dependence of the diffusive enhancement of these heavy components and not to an attenuation effect during propagation or a source cutoff.\footnote{Note that  the CRs emitted at redshifts close to $z_i$ would have traveled almost 300~Mpc in this scenario, and hence the heavy nuclei with energies larger than $\sim 50$~EeV may be further suppressed with respect to what shown in fig.~\ref{fig:spcomp}, where for simplicity we neglect  their interactions with the background photons. This could actually even improve the fit to the data.}
The diffuse extragalactic flux from all the sources present up to high redshifts, for which we adopted the SFR evolution,  dominates the flux below $\sim 5$~EeV and down to $\sim 0.1$~EeV, where the transition to a dominant Galactic CR origin would take place. Note that the spectral slope $\gamma=2.7$ of the diffuse component may reflect the adopted common spectral slope of the sources that contribute to it although it may also result as an effective slope if many sources having harder spectra but with a distribution of cutoff energies are present \cite{ka06}. 

Let us also mention that a low-energy suppression of the diffuse extragalactic flux may appear due to a magnetic horizon effect if the closest sources contributing to it are not very nearby, so that due to the diffusion the low-energy particles do not manage to reach the observer even from the closest ones. This effect is however expected to be milder for the source evolution following the SFR, for which the bulk of the contribution at low energies comes from high redshifts, and hence for simplicity  we ignored it. Note also that, due to the `propagation theorem' \cite{propa}, the diffuse spectrum is not expected to be modified by the effects of magnetic fields at higher energies, and it should not depend on the density of sources as long as the interaction length of the CRs with the background photons remains much larger than the typical intersource separation. Since at the highest energies the diffuse component is anyhow negligible in the scenario considered, these attenuation effects would not be relevant.

The amount of secondary protons that get produced is subdominant in this scenario, peaking at about 0.5~EeV (when multiplied by $E^3$). 
Regarding the composition, which is inferred from the measured depth of maximum of the air showers, $X_{\rm max}$, the main trends observed are reproduced, with $\langle {\rm ln}A\rangle$ decreasing slowly up to $E\sim 2$--3~EeV and the average composition becoming heavier at larger energies. This change, which is related with the observed change in  the slope of the $X_{\rm max}$ dependence with energy (i.e. the `elongation rate'), is accounted here by the fast increase of the N component arriving from the nearby source. This leads to a change in the slope of $\langle {\rm ln}A\rangle$ slightly below the energy of the ankle, as is indeed observed. The composition results depicted are those inferred from the $X_{\rm max}$ measurements adopting Sibyll~2.3 as the model for the  hadronic interactions, and the data points of  $\langle {\rm ln}A\rangle$ would shift slightly downwards for the EPOS-LHC or QGSJET~II-04 models \cite{augerxm}. 

Regarding the variance, $\sigma^2({\rm ln}A)$, it decreases steadily for increasing energies, showing a more pronounced drop near the ankle energy and remaining then quite small. The data points show a similar trend, although they are systematically below the model expectations. One has to keep in mind however that some of the experimentally inferred points for $\sigma^2({\rm ln}A)$ actually have negative central values (even more so for other hadronic models), which is unphysical and may be pointing towards some issues with the hadronic models.

\begin{figure}[t!]
\centering
\includegraphics[scale=.9,angle=0]{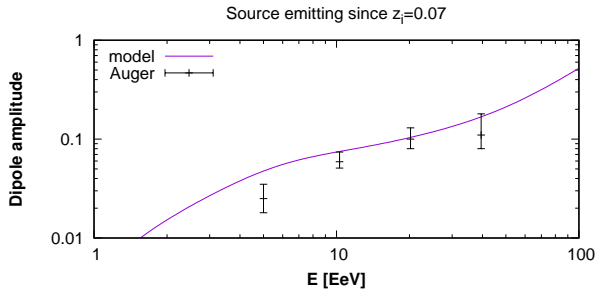}
\caption{Dipole amplitude as a function of the energy resulting from the extragalactic nearby source scenario of fig.~\ref{fig:spcomp}, compared with the amplitudes determined by the Pierre Auger Observatory \cite{dipoleapj} in the bins [4, 8]~EeV, [8, 16]~EeV, [16, 32]~EeV and $E>32$~EeV.}
\label{fig:dipole}
\end{figure}

In fig.~\ref{fig:dipole} we show the amplitude of the dipolar anisotropy that is produced by the nearby source, assuming that the diffuse flux is isotropic. We also ignore the contribution to the anisotropy from the Galactic component, which should be small above 1~EeV  due to the relatively small fraction of cosmic rays of Galactic origin at these energies, even though the intrinsic anisotropy of this component alone would be large. The overall agreement with the dipolar amplitude measured at different energies by the Auger Collaboration \cite{dipoleapj} is quite reasonable, especially if one keeps in mind that the Galactic magnetic field effects (neglected here) should reduce the amplitude of an extragalactic dipolar cosmic-ray distribution that is observed from the Earth, besides changing its overall direction \cite{ha10}. For instance,  due to the deflections in the regular Galactic magnetic field the extragalactic  dipolar amplitude observed at the Earth would typically be within 40\% to 100\% of its original value for rigidities $E/Z=5$~EeV, and would be between 20\% and 90\% of its original value for $E/Z=2$~EeV. The actual value of the suppression depends on the original direction of the dipolar anisotropy. In particular, for a dipole in the direction of Cen~A and adopting the Galactic magnetic field model from \cite{jf12}, the dipole amplitude would be suppressed by a factor of about 0.85 for the two rigidity values mentioned above.

\begin{figure}[t!]
\centering
\includegraphics[scale=1.3,angle=0]{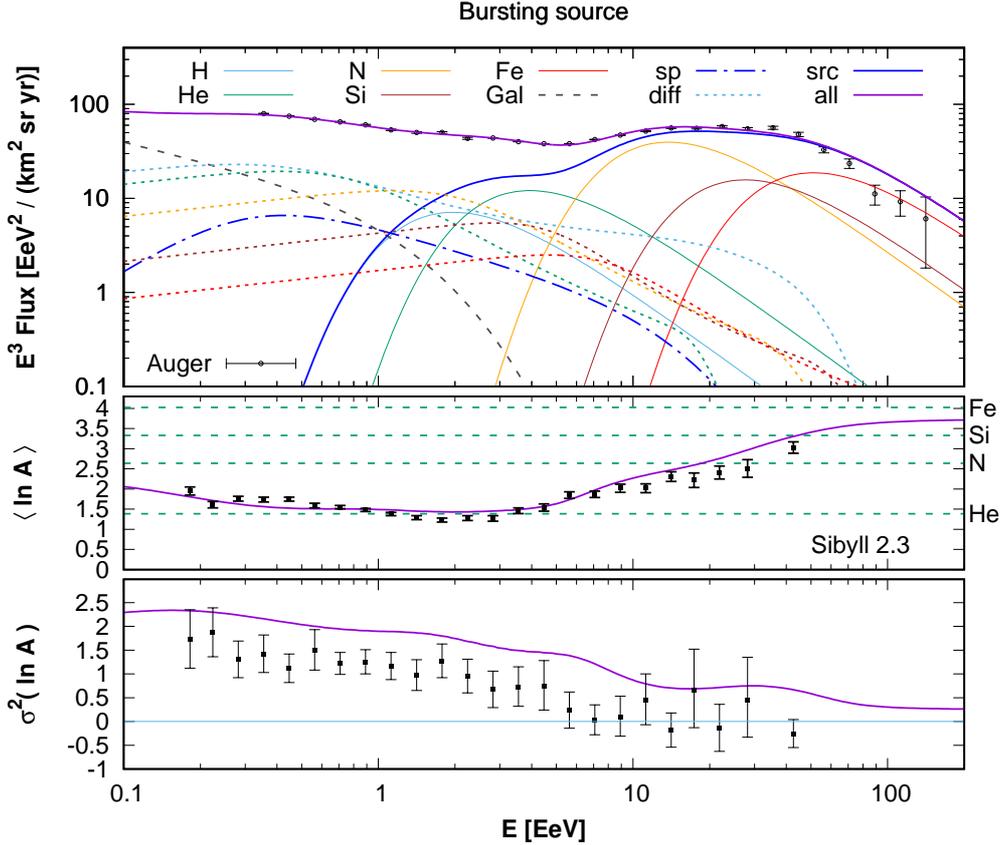}
\caption{ Energy spectrum (top panel), average logarithmic mass (middle panel) and variance of ln$A$ (bottom panel) vs. energy for the model with the bursting source, with the parameters described in the text, compared with data from the Pierre Auger Observatory \cite{augersp17,augerxm}.}
\label{fig:spcompb}
\end{figure}

\begin{figure}[t!]
\centering
\includegraphics[scale=.9,angle=0]{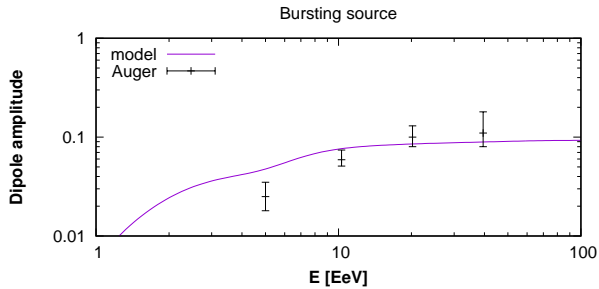}
\caption{Dipole amplitude as a function of the energy resulting from the nearby extragalactic source scenario of fig.~\ref{fig:spcompb}, compared with the determinations by the Pierre Auger Observatory \cite{dipoleapj}. }
\label{fig:dipoleb}
\end{figure}

In figures~\ref{fig:spcompb} and \ref{fig:dipoleb} we show the spectrum, composition and anisotropy results for a scenario with a source having a burst of emission at redshift $z_b$. We also consider $r_{\rm s}=4$~Mpc and adopt $z_b=0.015$ so that, for the energies at which the nearby source dominates the flux, its dipolar anisotropy is at the 10\% level. We take $B=50$~nG and $l_{\rm c}=0.05$~Mpc, which lead to $E_{\rm c}\simeq 2.2$~EeV. Due to the fact that  in the bursting scenarios the shape of the enhancement factor beyond its maximum is steeper than in the previous case, we adopt in this case a harder source flux with  $\gamma_s=2$. The fractions considered are 
$f^s_{\rm H}=0.34$, $f^s_{\rm He}=0.30$, $f^s_{\rm N}=0.27$, $f^s_{\rm Si}=0.05$ and $f^s_{\rm Fe}=0.04$. For the diffuse fluxes we adopt the same parameters as in the previous example. One can see that the overall agreement with the data is quite good also in this case. Note that for the bursting scenario the dipolar amplitude from the nearby source is expected to be independent of the energy, and the change in the amplitude observed in fig.~\ref{fig:dipoleb} actually arises from the energy dependence of the fractional contribution to the total flux that is due to the nearby source. 
The detailed energy dependence of the dipole anisotropy may then help to discriminate between the different scenarios discussed in this work, in which a nearby source dominates the flux above few EeV, and may eventually  also discriminate these models from those that consider a large number of sources that follow the overall distribution of galaxies, as discussed in \cite{hmrheavy,ha18,di18,gl19}.

\section{Summary}
We have studied in detail the spectrum that would be observed from a CR source in the diffusive regime. A strong enhancement of the flux appears at energies below the transition from the rectilinear to the diffusive regimes, i.e. for $E<E_{\rm rect}=\sqrt{r_{\rm s}/l_{\rm c}}E_{\rm c}$. For a steady source, the enhancement factor scales as $E^{-2}$ for $0.2E_{\rm c}\ll E<E_{\rm rect}$ and as $E^{-1/3}$ (for Kolmogorov turbulence) for $E\ll 0.1E_{\rm c}$.  Taking into account the finite age of the source, a  suppression appears in the spectrum at energies below the value for which the diffusion time from the source to the observer becomes comparable to the age of the source, i.e. for $l_D< r_{\rm s}^2/d(z_i)$. 

Using these results, we proposed a possible  scenario to explain the different UHECR observations in which the bulk of the CRs above  few EeV come from a relatively nearby source inside the Local Supercluster volume while at lower energies and down to  $\sim 0.1$~EeV the majority of the observed CRs result from the cumulative contribution from all the other extragalactic sources present up to high redshifts. It is important to keep in mind that, in order that the flux be enhanced, the nearby source needs to be within the Local Supercluster region in which the large magnetic fields lead to diffusion up to ultrahigh energies. On the other hand, the cumulative fluxes from the CR  sources outside this region are not expected to be strongly enhanced by the diffusion. In particular, a steady isotropic extragalactic flux originating outside the region of strong magnetic fields should remain isotropic to the observer (as implied by Liouville's theorem) and should not develop any density gradient inside the Local Supercluster region (as required by the diffusion equations in the absence of sources).

For illustration, we considered a scenario in which a nearby source at 4~Mpc distance is emitting since a redshift $z_i\simeq 0.07$ and another  one in which the source had a burst at a redshift $z_b=0.015$. We adopted a magnetic field in the Local Supercluster such that $E_{\rm c}\simeq 2$--3~EeV. The composition and spectral index of the source and those of the diffuse extragalactic component were taken so as to approximately reproduce the observations,
and, indeed, these simplified scenarios can account for all the main features of the spectrum, composition and large-scale anisotropies measured at ultrahigh energies. 
One should keep in mind that the specific values inferred for all those parameters will ultimately depend on the assumptions about the evolution of the sources that give rise to the diffuse component, the density of those sources, the actual time dependence of the nearby source emissivity, the adopted spectral distribution of the turbulence of the magnetic field, its strength, coherence length and possible nonhomogeneities in its distribution or any other departures from the idealized scenario considered. They would also change depending on the hadronic model that is used to interpret the composition measurements.

Note that given the energy dependence of the effects that modulate the spectrum of the nearby source, the required CR spectral slopes at the sources can be well compatible with the expectations from diffusive shock acceleration, even allowing for some steepening due to inefficiencies in the acceleration process. 

The flux from the nearby source is expected to show a dipolar anisotropy having an increasing amplitude as a function of energy, with typical values that are compatible with those observed by the Pierre Auger Observatory \cite{dipolescience,dipoleapj}.
At energies above $\sim 10$~EeV the two scenarios considered  lead however to different predictions, with the amplitude of the anisotropy remaining quite  flat for the bursting source case while it would keep increasing if the source emission continued up to more recent times. 
One should also expect that at the highest energies, as the surviving CR components  tend towards the quasirectilinear regime, more localized anisotropies, on scales of few tens of degrees, could start to become observable. Indeed, already some hints for this kind of signatures have been reported above 40~EeV \cite{apj2015,hotspot,tahotspot}. Although we considered for simplicity a scenario with just a single dominant nearby source, the possibility that several nearby sources contribute at
ultrahigh energies  clearly exists. The study of the localized anisotropies appearing at the highest energies can be helpful to know if there is more than one nearby source and to identify where the sources are located.

\section*{Acknowledgments} We thank D. Harari for useful discussions. This work was supported by CONICET (PIP 2015-0369) and ANPCyT (PICT 2016-0660). We thank the Auger Collaboration for making data available at www.auger.org and to M. Unger and J. Bellido for help.

\end{document}